\documentclass[12pt]{article}
\usepackage{natbib}
\usepackage{amsmath}
\usepackage{bm,booktabs,color,xcolor,siunitx}
\usepackage{amssymb}
\usepackage{amsfonts}
\usepackage{multirow}
\usepackage{slashbox}
\RequirePackage[OT1]{fontenc}
\RequirePackage[colorlinks,citecolor=blue,urlcolor=blue]{hyperref}
\usepackage[ruled,vlined,linesnumbered]{algorithm2e}
\usepackage{caption}
\usepackage{subcaption}
\usepackage{graphicx,psfrag,epsf}
\usepackage{enumerate}

\definecolor{lightgray}{gray}{0.9}
\newtheorem{theorem}{Theorem}
\newtheorem{proposition}{Proposition}
\newtheorem{proof}{Proof}
\newtheorem{example}{Example}
\newtheorem{remark}{Remark}
\newcommand{\pib}{\mbox{\boldmath{$\bf \pi$}}}
\newcommand{\xib}{\mbox{\boldmath{$\bf \xi$}}}
\newcommand{\NN}{\mbox{\boldmath{$\bf N$}}}
\newcommand{\CC}{\mbox{\boldmath{$\bf C$}}}
\newcommand{\1}{\mbox{\boldmath{$\bf 1$}}}
\newcommand{\0}{\mbox{\boldmath{$\bf  0$}}}
\newcommand{\M}{\mbox{\boldmath{$\bf M$}}}
\begin{document}
%
%
%\centerline{EXAMPLE}
%\vskip 3mm

\noindent Distribution-Free Control Charts Based on Runs and Patterns
\vskip 3mm

\vskip 5mm
\noindent Tung-Lung Wu

\noindent Department of Mathematics and Statistics

\noindent Mississippi State University

\noindent Mississippi state, MS \ 39762

\noindent tw1475@msstate.edu

\vskip 3mm

\noindent ABSTRACT

Exact distribution-free runs and patterns-type control charts for monitoring the unknown target value (or the unknown location parameter) are proposed for both continuous and discrete data. The in-control run length distributions can be obtained through a combination of random permutation and conditioning arguments. Data-dependent control limits can be determined by the finite Markov chain imbedding technique, which is commonly used to compute the conditional distributions of runs and patterns.  Two common runs-type statistics, the longest run and  scan statistics, are studied in detail to illustrate the method.  Numerical results and an application to piston rings data are given to evaluate the performance of the proposed control charts. 
\vskip 4mm

\noindent 1.   Introduction and Notations

Statistical process control (SPC) is an effective technique to monitor the  characteristics of a process. 
The purpose of a control chart is to detect any process changes, such as location and scale shifts, as quickly as possible. The typical setting is as follows:
suppose there are $m_0$ historical independent and identically distributed (i.i.d.) data $X_{-m+1},\ldots,X_0$, and the future data are collected according to the following model~\citep[see for example,][]{Zhou-2009}
\begin{align}
X_t  &= \left\{\begin{array}{lll} \label{test1}
% 0 & \mbox{if} \,X_t-X_{t-1}>=2,\\
F_0(x;\mu_0) & \mbox{for } t=-m+1,\ldots,0,1,\ldots,\tau  ,\\
F_1(x;\mu_1)   & \mbox{for } t=\tau+1,\ldots,
\end{array}\right. 
\end{align} 
where $\tau$ is the unknown change point. 
In a typical control chart, the plotted data points are assumed to follow a parametric distribution $F_0$, such as normal distribution. The problem is clear; if the distributional assumption is not satisfied, the promised characteristics of the control chart, such as its in-control average run length  (ARL), can no longer be reliable. In other words, such control charts are not robust to the underlying distributions. Recent applications show the underlying distribution might  not be normal. There is a need for new control charts with no distributional assumptions.   Many nonparametric or distribution-free control charts have been proposed for the purpose of controlling  the desired in-control ARL. Hence the main objective of this manuscript is to develop distribution-free control charts for detecting location shifts. 

In the past decade, nonparametric or distribution-free control charts have received considerable attention in the area of quality and process control. 
The main stream of nonparametric or distribution-free control charts consists of rank-based control charts and likelihood-based control charts. There are generally three types of control charts: exponentially weighted moving average (EWMA), cumulative sum (CUSUM), and Shewhart control charts. The conventional Shewhart control charts can be enhanced with supplemented runs rules. EWMA and CUSUM are popular because they are superior in detecting small mean shifts. Examples of nonparametric EWMA control charts can be found in the works of \cite{Amin-1991}, and \cite{Zou-2010}. Nonparametric CUSUM control charts have been   studied by \cite{Chatterjee-2009}, \cite{Chowdhury-2015}, and \cite{Li2013}.  A comprehensive review of nonparametric  control charts is given by \cite{Chakraborti-2001}. 

The simplicity, easy implementation and computation of its run-length characteristics make the Shewhart-type charts preferable.   Runs rules have also been used to enhance the Shewhart control chart for small mean shifts. Runs-type control charts are used to increase Shewhart control charts' capacities for detecting small mean shifts, as they have done for scan statistics \citep[see for example,][]{Shmueli-2003}. However, the applicability is limited and the performance is not guaranteed if normality or some parametric models are assumed   \citep[see for example,][]{Charles-1987,koutras-2007}. To relax this restriction, \cite{Chakraborti-2009} proposed a Phase-II nonparametric control chart based on precedence statistics with runs rules for monitoring the unknown location parameter. Their charts signaled when two consecutive plotting points, such as the median, fall outside the control limits. Another result based on Wilcoxon signed-rank statistics with some runs-type rules is given by \cite{Chakraborti-2007}. In contrast to the  chart proposed by \cite{Chakraborti-2009}, the Wilcoxon signed-rank statistic is applied under the known median case. \cite{Balakrishnan-2010} present a general class of nonparametric control charts based on order statistics and provide a detailed review of the area.

Nonparametric Shewhart control charts with runs rules seem to be underdeveloped relative to nonparametric EWMA and CUSUM control charts. A lack of methods for handling general runs without any distributional assumptions is one likely cause.  

The finite Markov chain imbedding (FMCI) technique has been used to evaluate the exact characteristics of the run length distribution (such as the mean and standard deviation) of  various control charts, including Shewhart, EWMA and CUSUM control charts where the normality is assumed \citep[see for example,][]{Fu-2002,Fu-2003}. Distribution-free runs and patterns-type control are discussed in here, with conditional distributions of run and patterns obtained exactly by the FCMI technique. Bernoulli trials are viewed given the total number of successes as random permutations. The idea of using conditional runs and patterns coincides with other work  in which the conditional distributions of the proposed monitoring statistics are independent of the  unknown parameters and/or the underlying distributions \citep[see for example,][]{Chen-2016}.

%Three classic control charts are Shewhart,EWMA  and  CUSUM control charts.  It is well-known that Shewhart control charts can quickly detect large shits while the EWMA and CUSUM control chart are better for small shifts.
%In practice, we prefer the Shewhart control chart since it is easy to understand and interpret by practitioners. Many supplemented rules are developed to improve the ability of the Shewhart control charts in detecting small shifts. Some well-known runs statistics are the longest run and scan statistic. In this manuscript, we will construct distribution-free charting procedures for these two runs statistics and study their properties. 

{Much of the literature for distribution-free runs-type control charts is either restricted to certain runs statistics or not exactly distribution-free. Our contribution is to 
construct a general framework for distribution-free runs and patterns-type control charts for both continuous and discrete data. In addition, the target value can be unknown. It is worth mentioning  the proposed control charts are truly distribution-free. The in-control ARL can be achieved exactly through a combination of random permutation and conditioning arguments, and the proposed charting procedures lead to data-dependent control limits. With some modification, the FMCI technique is used to compute the necessary conditional probabilities that are used to maintain the in-control ARL. See Appendix for more details.
}
 
The rest of this manuscript is organized as follows. Section 2 describes the conditional distributions of scan statistics and the longest run. Section 3 
provides the charting procedures for  scan-based and runs-based control charts.
Section 4 provides the numerical results.  Section 5 studies an application to piston rings data. Section 6 provides the summary and discussion.
\vskip 3mm

\noindent 2. Conditional Distributions

Details of the finite Markov chain imbedding technique are given for computing  the exact conditional distributions of scan statistics and the longest run in this section.

\vskip 3mm

\noindent 2.1 Scan Statistics

 Let $X_1,\ldots,X_n$ be a sequence of i.i.d. Bernoulli trials with $p=P(X_1=1)$.  The scan statistic is defined as 
 \begin{equation}
 	S_n(r) = \max_{1 \leq t\leq n-r+1} S_n(r,t),
 \end{equation}
 where $S_n(r,t)=  \sum_{i=t}^{t+r-1}X_i$, and $r$ is the window size. \cite{Fu-Lou-Wu-2012} used 
 the FMCI technique to obtain the exact conditional distributions of scan statistics. %A brief description of the FCMI technique is given below in preparation for the conditional distribution of the longest run in the next section.
 
Considering a sequence of Bernoulli trials with the two possible outcomes of $\{0,1\}$ and $S_2=\{0,1\}$, let $\Lambda = \cup^{L}_{i=1} \Lambda_i$ be a compound pattern consisting of $L$ simple patterns $\Lambda_1,\Lambda_2,\ldots,\Lambda_L$, where each simple pattern $\Lambda_i$ is composed of a specified sequence of 2 symbols $\{0,1\}$, and the length  of $\Lambda_i$ is fixed. Denote the waiting time of the first occurrence of the compound pattern $\Lambda$ in a sequence by $W(\Lambda)$. The distribution of a scan statistic is calculated based on the FMCI technique through the dual relationship with the waiting time distribution of an associated compound pattern. An example is given below.
 
 \begin{example} Let $r=5$ and $s=2$. The compound pattern associated with the event $\{S_n(5)<2\}$ is $\Lambda_{5,2}=\{11,101,1001,10001\}$. Thus, the scan statistic probability is viewed as the probability that the pattern will not occur in the sequence of length $n$, and we have  $P(S_n(5)<2) = P(W(\Lambda_{5,2})>n).$
 	In the conditional case, the dual relationship still holds, i.e.
 	\begin{align}
 		&P\left(S_n(5)<2\Big|\sum^{n}_{i=1} X_i=m\right) = P\left(W(\Lambda_{5,2})>n\Big|\sum^{n}_{i=1} X_i=m\right).
 	\end{align} 
 \end{example} 
 
 Given $r$ and $s$, a set of example patterns of lengths no longer than $r$ is defined as
 $$\Lambda_{r,s} = \{\Lambda_i,i=1,\ldots,\ell \},$$ 
 where $\Lambda_i$ is a simple pattern that begins and ends with $1$  and contains total $s$ 1's, and $\ell$ is the total number of simple patterns corresponding to the scan statistic with parameters $r$ and $s$. For each scan statistic probability $P(S_n(r)<s)$, the total number of simple patterns is
 $$\ell = 
 \sum^{r-s}_{\nu=0}\binom{s-2+\nu}{\nu}.$$
 Let $$\mathcal{P} = \left\{\pib=(\pi_1,\ldots,\pi_n): \pi_i=0,1 \mbox{ and } 	\sum^{n}_{i=1}\pi_i=m \right\}$$
 be the family of random permutations with $m$ 1's and $n-m$ 0's. Then the conditional distributions of runs and patterns, given the total number $m$ of successes for a sequence of $n$ Bernoulli trials, are the same as the distributions of runs and patterns in an $[n-m,m]$-specified random permutation $\pib=(\pi_1,\ldots,\pi_n)$. It is notable that in an $[n-m,m]$-specified random permutation, the distributions of runs and patterns are independent of $p$. 
 
 To construct a finite Markov chain $\{Y_t\}$,  $E_{r,s}$ is defined as a set of all subpatterns of $\Lambda_{r,s}$. An imbedded Markov chain can be defined on the state space   
 $$\Omega = \{(l,\omega): l = 0,1,
 \ldots,m \mbox{ and } \omega \in E_{r,s}\cup S_2\}\cup\{\emptyset,\alpha\},$$
 where $\emptyset$ is the initial state and $\alpha$ is the absorbing state, meaning that the compound pattern $\Lambda_{s,r}$ occurs if the chain enters the absorbing state. Let $\ell_m+1$ denote the size of the state space $\Omega$. The  transition probabilities from state $u=(m_{t-1},\omega_{t-1})$ to  state $v=(m_t,\omega_{t})$ are 
 \begin{align}
 	p_{uv}(t) &= P(Y_t = (m_{t},\omega_{t})|Y_{t-1}=(m_{t-1},\omega_{t-1})),\nonumber\\
 	&= \left\{\begin{array}{lll}\label{tran}
 		\frac{m-m_{t-1}}{n-t+1} & \mbox{if} & \pi_t=1, m_t=m_{t-1}+1 \mbox{ and } \omega_t=<\omega_{t-1},1>_{E_{r,s}},\\
 		\frac{n-m-t+m_{t-1}+1}{n-t+1} & \mbox{if} & \pi_t=0, m_t=m_{t-1} \mbox{ and } \omega_t=<\omega_{t-1},0>_{E_{r,s}},\\
 		1 & \mbox{if} & \omega_t =\omega_{t-1}= \alpha,\\
 		0 & \mbox{if} & \mbox{otherwise},\\
 	\end{array}\right. 
 \end{align}
 where, after $\pi_t$ is observed, $<\omega_{t-1},\pi_t>_{E_{r,s}}$ denotes the longest subpattern. Therefore, the transition matrices are of the form
 \begin{equation}\label{Eq5}
 	\M^1_t(m) =\left[ \begin{tabular}{c|c}
 		$ \NN^1_t(m)$  & $ \CC^1_t(m)$ \\ \hline
 		$ \0$  &  $1$
 	\end{tabular} \right]_{(\ell_m+1)\times(\ell_m+1)}, %\label{eq:2.1}
 \end{equation}
 where $\NN^1_t(m),t=1,2,\ldots,$ are $\ell_m\times\ell_m$ matrices.

 The subsequent theorem describes the exact conditional distribution of $S_n(r)$.

 \begin{theorem} \label{thm1}
 	Let $X_1,\ldots,X_n$ be a sequence of i.i.d. Bernoulli trials. Then 
 	\begin{align}\label{cond}
 		P\left(S_n(r)<s\Big|\sum^{n}_{i=1} X_i=m\right)=P\left(W(\Lambda_{r,s})>n\Big|\sum^{n}_{i=1} X_i=m\right) = \xib_0\prod^n_{t=1}\NN^1_t(m)\1^{\top},
 	\end{align}
 	where $\xib_0$ is a $1\times\ell_m$ vector of initial probabilities, $\NN^1_t(m),t=1,\ldots,n,$ are $\ell_m\times\ell_m$ matrices whose entries are given in (\ref{tran}), and $\1$ is a $\ell_m\times1$ row vector of ones.  
 \end{theorem}
 \begin{proof}
 	Based on the above construction, the conditional waiting time variable $W(\Lambda_{r,s})$ given the total number of successes is finite Markov chain imbeddable and it follows from Theorem 2.1 of \cite{Fu-lou-2003} that the exact distribution is of the form in (\ref{cond}).
\end{proof}
 
 The conditional distribution in (\ref{cond}) is independent of $p$.  \cite{Fu-Lou-Wu-2012} provide more details of the imbedded Markov chain.
 
 \vskip 3mm
 
2.2 The Longest Run

 Let $L_n$ denote the length of the longest  run of 1 in a sequence of Bernoulli trials, and $S_2=\{0,1\}$. Consider the run $$\Lambda_{d} = \underbrace{1\cdots1}_{d}.$$
 The event $\{L_n<d\}$ occurs only if the run $\Lambda_{d}$ does not appear in the sequence of Bernoulli trials. Thus,  
 \begin{align}
 	P(L_n<d) = P(W(\Lambda_{d})>n).
 \end{align}
 The above equation persists for conditional distribution given the total number of successes. 
 
 The previous section describes the construct of the imbedded Markov chain for scan statistics. Similarly, a set of subpatterns can be defined as
 $$E_d = \{1,11,\ldots,\underbrace{1\cdots1}_{d-1}\}. $$
 Thus a finite nonhomogeneous Markov chain $\{Y_t\}$ can be defined on the state space 
 \begin{align}
 	\label{L_omega}
 	\Omega = \{(l,\omega): l = 0,1,
 	\ldots,m \mbox{ and } \omega \in E_{d}\cup S_2\}\cup\{\emptyset,\alpha\},\end{align}
 and the transition probabilities from state $u=(m_{t-1},\omega_{t-1})$ to state $v=(m_t,\omega_{t})$ are 
 \begin{align}
 	p_{uv}(t) &= P(Y_t = (m_{t},\omega_{t})|Y_{t-1}=(m_{t-1},\omega_{t-1})),\nonumber\\
 	&= \left\{\begin{array}{lll}
 		\frac{m-m_{t-1}}{n-t+1} & \mbox{if} & \pi_t=1, m_t=m_{t-1}+1 \mbox{ and } \omega_t=<\omega_{t-1},1>_{E_{d}},\\
 		\frac{n-m-t+m_{t-1}+1}{n-t+1} & \mbox{if} & \pi_t=0, m_t=m_{t-1} \mbox{ and } \omega_t=<\omega_{t-1},0>_{E_{d}},\\
 		1 & \mbox{if} & \omega_t =\omega_{t-1}= \alpha,\\
 		0 & \mbox{if} & \mbox{otherwise}.\\
 	\end{array}\right. 
 \end{align}
 Again, the transition matrices are of the form
 \begin{equation}
 	\M^2_t(m) =\left[ \begin{tabular}{c|c}
 		$ \NN^2_t(m)$  & $ \CC^2_t(m)$ \\ \hline
 		$ \0$  &  $1$
 	\end{tabular} \right]. %\label{eq:2.1}
 \end{equation}
 The exact conditional distribution of $L_n$ is 
 $$P\left(L_n<d\Big|\sum^{n}_{i=1} X_i=m\right) = \xib_0\prod^n_{t=1}\NN^2_t(m)\1^{\top}.$$
 
 \cite{Lou-1996} describes studies of the conditional longest run.
 However, the procedure is different in \cite{Lou-1996} relative to the procedure we used to to produce the finite Markov chains.
 It can be clearly seen from this construction of the finite Markov chain that  the conditional distribution is independent of $p$; although the conditional distribution does not depend on $p$,  Lou (1996)'s formula  involves $p$.  Lou (1996)'s formula has the advantage of allowing the Bernoulli trials to be Markov dependent.

 \begin{example}
 	Let $n=5$, $m=3$, and $d = 3$. The pattern corresponding to $\{L_5<3\}$ is $\Lambda_{3}=111$. The ending block is $E_{3} = \{1,11\}$, and $E_{3}\cup S=\{0,1,11\}$.  	
 	The state space can be constructed according to (\ref{L_omega}) and is given by $\Omega$ = \{(0,0),(1,0),(1,1),(2,0),(2,1),(2,11),(3,0),(3,1), (3,11)\}$\cup$\{$\emptyset$, $\alpha$\}. Some redundant states are removed from the state space. For example, the state $(1,11)$ will never occur.
 	 Theorem~\ref{thm1} gives a probability of 0.7 that the length of the longest  run is less than 3. The same probability can be obtained by enumeration. Given 3 successes (1's) in 5 Bernoulli trials, there are 10 possible outcomes \{11100,
 	11010,10110, 01110,11001,10101,01101,10011,01011,00111\}. Seven outcomes \{11010,10110,
 	11001,10101, 01101,10011,01011\}  contain the longest run of length less than 3,  hence the probability is 7/10. 
 \end{example}

 \begin{remark}
 	The two statistics, the scan statistic and the longest run, are used for illustrative purposes. In fact, Theorem~\ref{thm1} can be use to compute the distributions of general runs and patterns of fixed lengths. The construction of the matrix $N_t$ depends on the runs or patterns used. 
 \end{remark}

\noindent 3.  Distribution-Free Control Charts

 In this section, new distribution-free runs-type control charts that function with any underlying process distribution $F_0$ are proposed. The methodology for one-sided control charts, as we describe them throughout this paper, are applied to detecting upward shifts, but this methodology can also be applied to control charts with lower and upper limits for detecting downward and upward shifts.

 Let $Y_1,\ldots,Y_n$ be the sequential observations from an unknown, underlying distribution $F_0$. 
 In the one-sided control chart, a label ``1'' is assigned to an observation if its value is greater than a certain threshold; otherwise, ``0'' is assigned. For example, consider
 \begin{align}\label{111}
 	X_n= \left\{\begin{array}{lll}
 		1 & \mbox{if} & Y_n\geq c,\\
 		0 & \mbox{if} &Y_n< c.\\
 	\end{array}\right. 
 \end{align}
 The longest run and scan statistic can then be defined for the  sequence $\{X_n\}$ with two outcomes.  It has been widely studied that runs and scan rules can enhance the ability of a Shewhart-type control chart for detecting small shifts. When $r=1$, the scan-based control chart is equivalent to the conventional Shewhart control chart, which is good for detecting large shifts. On the contrary, one can use a large $r$ with a smaller threshold $c$ to detect small shifts. This idea has been verified by our numerical study. 
 A general runs and patterns rules, denoted by $R(k,r,Z)$, is proposed by \cite{Shmueli-2003}. The rules in these notations are described below: if the $k$ of the last $r$-tested points fall in the region $Z$,  the control chart signals an out-of-control alert.

 %The eight zones which partition the real line are given below:
 %\begin{itemize}%
 %	\item[] $Z_{(3,\infty)}$ = the interval (3,$\infty$)
 %	\item[]$Z_{(2,3)}$ = the interval (2,3)
 %	\item[] $Z_{(1,2)}$ = the interval (1,2)
 %	\item[] $Z_{(0,1)}$ = the interval (0,1)
 %	\item[] $Z_{(-1,0)}$ = the interval (-1,0)
 %	\item[] $Z_{(-1,-2)}$ = the interval (-2,-1)
 %	\item[] $Z_{(-2,-3)}$ = the interval (-3,-2)
 %	\item[] $Z_{(-\infty,-3)}$ = the interval ($-\infty$,-3).
 %\end{itemize}
Two types of rules are considered:
 \begin{enumerate} 
 	\item[(1)] R-1 = $R(k,r,(c,\infty))$: scan rule 
 	\item[(2)] R-2 = $R(k,k,(c,\infty))$: runs rule
 \end{enumerate}
 In rule (1), it is clear $k<r$ . Let $N_n = \sum_{i=1}^{n}X_i$. 
 %Given the total number of 1's at time $n$, a distribution-free scan-based control chart with data-dependent control limits can be constructed as follows.
 Given a pre-specified $\alpha$, the charting procedure for the distribution-free scan rule (R-1) control chart is given below.  
 %\RestyleAlgo{boxruled}
 %\LinesNumbered
 \begin{algorithm}[!h]
 	\caption{The scan rule control chart (R-1)}
 	\begin{enumerate}
 		\item Given a pre-specified $\alpha$, the control limits $\{c_n(\alpha)\}$   are determined by the \\ following equations: 
 		\begin{align}
 			&\nonumber	P(S_n(r)\geq c_n(\alpha)|N_n)\leq \alpha, \mbox{ for } n=1,\\ 
 			&\label{pro1}	P(S_n(r)\geq c_{n}(\alpha)|S_{n-1}(r)<c_{n-1}(\alpha),N_{n})\leq \alpha, \mbox{ for } n>1.   
 		\end{align} 
 		\item Based on $\{c_n(\alpha)\}$ found in step 1, the run length  is given by 
 		\begin{equation}
 			RL = \min\{n: S_n(r)\geq c_n(\alpha), n\geq 1\}.
 		\end{equation}
 	\end{enumerate}
 \end{algorithm}
 %Note that the monitoring process starting at time $\nu$ is because that the probability $P(S_n(r)\geq c_n(\alpha)|N_n)$

 Similarly, the charting procedure for the distribution-free  control chart with the longest run rule (R-2)  is given in Algorithm 2. 
 \begin{algorithm}[!h]\label{a1}
 	\caption{The longest run rule control chart (R-2)}
 	\begin{enumerate}
 		\item Given a pre-specified $\alpha$, the control limits $\{k_n(\alpha)\}$   are determined by the \\ following equations: 
 		\begin{align}
 			&\nonumber	P(L_n \geq k_n(\alpha)|N_n)\leq \alpha, \mbox{ for } n=1,\\ 
 			&	P(L_n\geq k_{n}(\alpha)|L_{n-1}<k_{n-1}(\alpha),N_{n})\leq \alpha, \mbox{ for } n>1.  \label{ln}
 		\end{align} 
 		\item Based on $\{k_n(\alpha)\}$ found in step 1, the run length  is given by 
 		\begin{equation}\label{longest}
 			RL = \min\{n: L_n\geq k_n(\alpha), n\geq 1\}.
 		\end{equation}
 	\end{enumerate}
 \end{algorithm}

These are non-trivial conditional probabilities.
 The detailed steps for computing the required conditional probabilities in (\ref{pro1}) and (\ref{ln}) are given in the Appendix. RL$_0$ is the in-control run length, RL$_1$ is the out-of-control run length, ARL$_0$ is the in-control average run length, and ARL$_1$ is the out-of-control average run length.
 %Note that in Algorithm~\ref{a1}, the probability $P(L_1\geq k_1(\alpha))$ is either 0 or 1 depending on the observed value of $N_1$. %Hence, it is suggested that the proposed longest run rule control chart may start monitoring a process after a small number of observations has been collected so that an appropriate $\alpha$ can be selected. However, we do not have this issue if there are $m_0$ reference observations available.   
 By controlling the conditional probabilities in (\ref{pro1}) and (\ref{ln}),  the RL$_0$ of our proposed control charts follows a geometric distribution. 
 \begin{proposition}
 	Suppose the process is in-control. We have $P(RL_0=n) = \alpha(1-\alpha)^{n-1}$ for $n\geq 1$ and for any underlying distribution.    
 \end{proposition}
 The proof is omitted here since it is similar to \cite{Chen-2016}'s proof in Theorem 1.
 Since the RL$_0$ distribution is the geometric distribution  with parameter $\alpha$, the ARL$_0$ is equal to $1/\alpha$. This result allows us to determine the exact ARL$_0$ through the conditional probabilities in (\ref{pro1}) and (\ref{ln}).
 
 \vskip 3mm
 \noindent 3.1 {Choice of $r$ and $c$}
 
 In this section, we suggest the appropriate combinations of the parameters $r$ and $c$ for detecting either small or large shifts. An unconditional scan-based control chart uses the scan rule to enhance the control chart for small shifts, and---when $r=1$---the control chart reverts  to the conventional Shewhart control chart. The combination of  a small $r$ and a large $c$ is used to detect a large shift and the combination of a large $r$ and a small $c$ is used to detect a small shift. 
 
 Although the threshold $c$ is not explicitly involved in the charting procedures, it plays an important role in the performance of the control charts. If a small $c$ is selected, there will be a large number of 1's . In this case, the true signal might be masked by noise. On the contrary, if a large $c$ is selected, a small number of 1's can be expected. In this case, the true signal might not be picked up due to the high threshold. Thus, the suggestion is to use a small $c$ for small shifts and a large $c$ for large shifts. 
 If there is information about some summary statistics, such as sample mean and sample standard deviation, use the conventional multiple standard deviation threshold. For a small shift, one or two standard deviations are suitable, while three standard deviations may be chosen for a large shift.
\vskip 3mm

\noindent 4. Numerical Results

To evaluate the performance of the proposed control charts in Section 3, $\alpha= 0.005$ are set, yielding the ARL$_0$ = $200$.  %under three family of distributions, normal and the $t$ and the gamma distributions.
 For ARLs, in each simulation 1000 sequences were generated from normal, $t$ and gamma distributions. 
 For robustness, we consider  $t$ and gamma distributions for the effects of heavy tails and skewness, respectively. The distributions were scaled to have a mean of 0 and a variance of 1. Therefore, $\mathcal{N}$(0,1), t(4) and gamma(1,1) are used in the simulations.  Let $\mu$ denote the difference between in-control and out-of-control process means. The knowledge of the target mean value is not needed. 
% and noncentral $t$ distributions with degrees of freedom 3 and 10. 
Distribution-free control charts usually assume there are $m_0$ historical (reference) data from the in-control process.  We select $m_0=20.$

\begin{figure}
	\centering
	\begin{minipage}{.5\textwidth}
		\centering
		\includegraphics[width=0.8\linewidth]{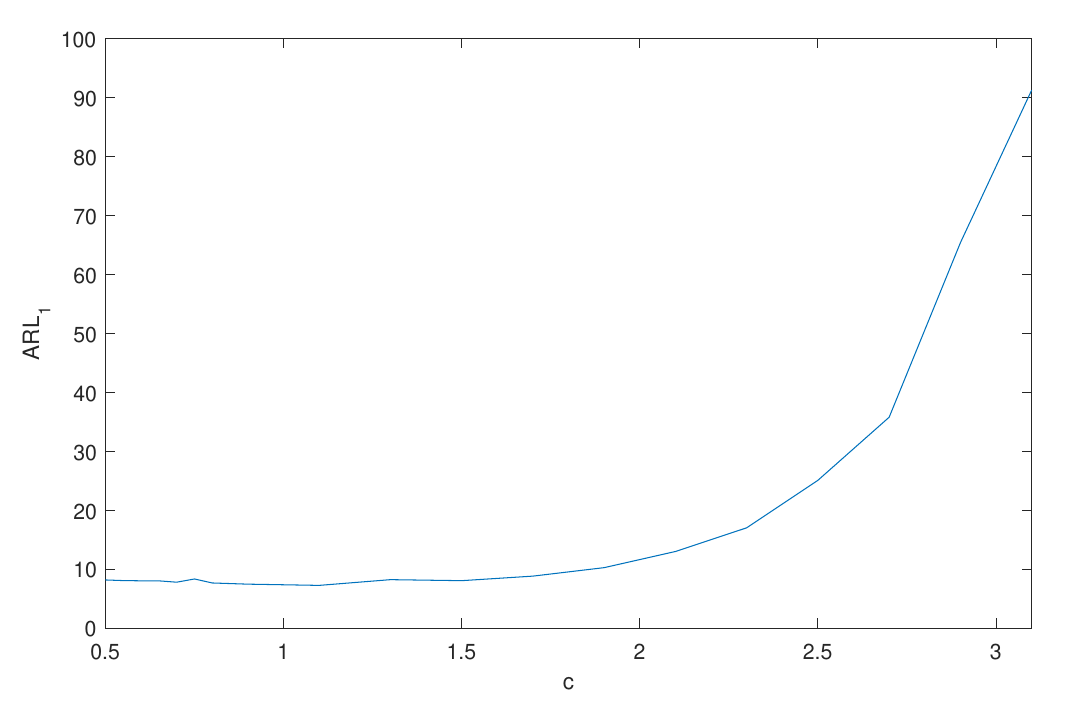}
		\captionof{figure}{ARL$_1$s of the R-2 control charts for $\mathcal{N}$(0,1) when $\mu=2$.}
		\label{f11}
	\end{minipage}%
	\begin{minipage}{.5\textwidth}
		\centering
		\includegraphics[width=0.8\linewidth]{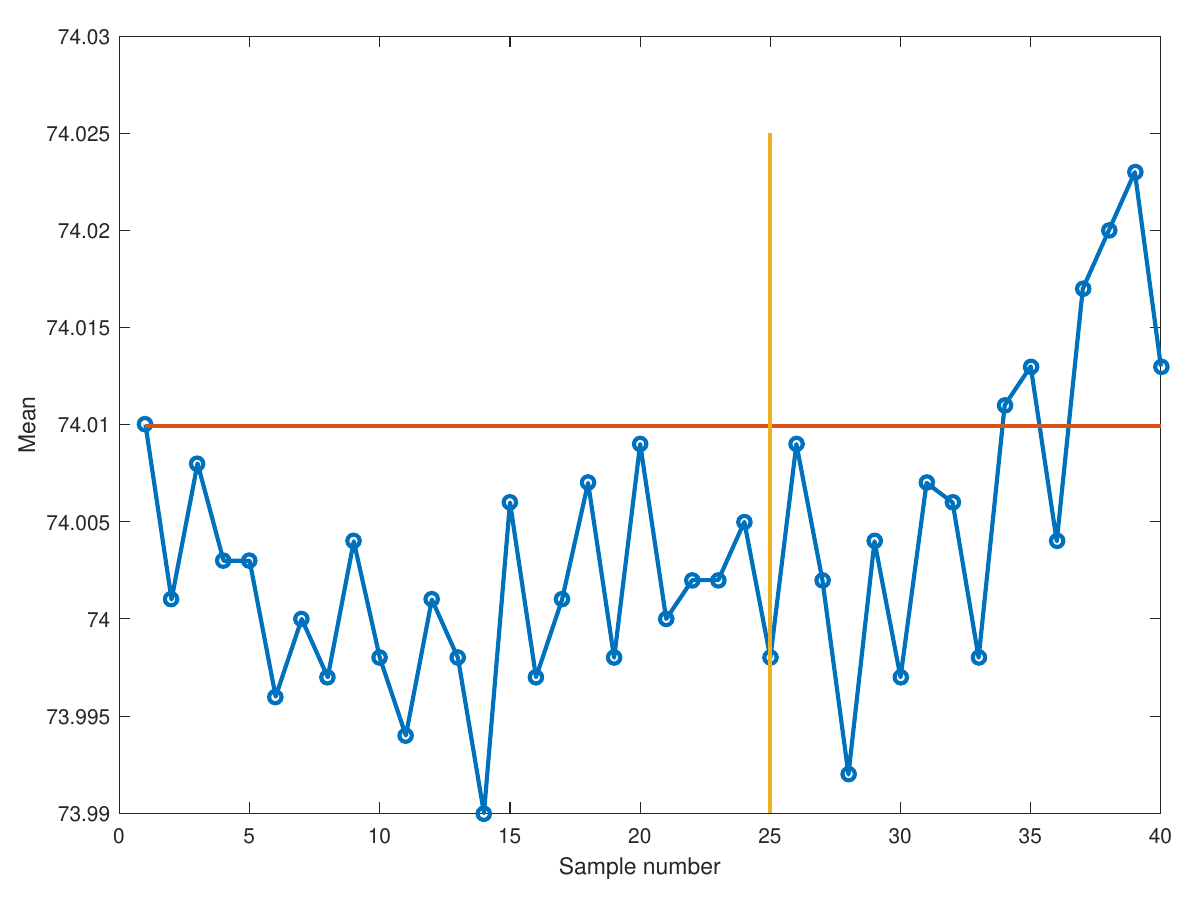}
		\captionof{figure}{The longest run rule control chart.}
		\label{dataplot}
	\end{minipage}
\end{figure}

Table~\ref{t-0} gives the empirical ARL$_0$s of the R-2 control charts for various mean shifts and values of $c$ under the normal, the t and  the gamma distributions. Under the null hypothesis that there is no mean shift, the ARL$_0$ stays the same regardless of the value of $c$; however, it largely drives the performance of the proposed control charts. Table~\ref{t-1} indicates that the values of $c$ producing the smallest ARL$_1$s for $\mu=1,2,$ and 3 are 0, 1, and 2, respectively. A guideline for choosing $c$ is to select a slightly smaller value than the anticipated mean shift $\mu$. 
The choice of $c=\mu-1$ for runs and patterns-type control charts is suggested. To examine this further, a plot of ARL$_1$s over a range of $c$ is given for the mean shift $\mu=2$.   
%a special case when $\mu=2$ and compare the ARL$_1$s for a certain range of $c$ is considered. 
In Figure~\ref{f11}, the smallest ARL$_1$ occurs at $c\approx1$, which supports our suggestion.
Hence, the ARL$_1$s of Table \ref{t-2} are only given for $\mu=1,2,3$ with $c=\mu-1$ under the $t$ and the gamma distributions. 

Table~\ref{t-11} gives the ARLs for the R-1 control charts. To ease the computational burden, $c=2$ was chosen. For scan-based control charts, the ARL$_1$s suggest  a larger window size and a smaller window size for small shifts ($\mu\leq 1$) and large shifts ($\mu\geq 3$), respectively. This coincides  with the choice of $r$ and $c$ given in Section~3.2. The ARL$_1$s for $\mu=1$ are large because the best $c$ for each value of $\mu$ was not selected.
The control charts work better when $F_1$ is stochastically larger than $F_0$. 
Table~\ref{T33} shows no relationship between the size of the reference data and the performance of the R-2 control chart.

\begin{table}[!tb]
	%\caption{Global caption}
	\caption{ARL$_0$s of the R-2 control charts for $\mathcal{N}$(0,1), $t$(4) and gamma(1,1).}
	\centering\label{t-0}
	
	\begin{tabular}{@{}|c|ccc|@{}}
		\hline
		%& \multicolumn{3}{c}{R-1 chart} & \multicolumn{3}{c}{R-2 chart}\\
		%	\cmidrule(lr){2-4} \cmidrule(l){5-7}
		&		$\mathcal{N}(0,1)$   &	$t$(4)  &	gamma(1,1)  \\\hline
		$c=1$&	225.24	 &	217.08	 &  225.82	\\
		$c=2$&221.71		 &	226.03	 & 	223.35	\\
		$c=3$&195.90		 &203.43		 & 	193.71	\\
		\hline	
	\end{tabular}

\end{table}

\begin{table}[!tb]
	%\caption{Global caption}
		\caption{ARL$_1$s of the R-2 control charts for $\mathcal{N}$(0,1).}
		\centering\label{t-1}
		\begin{tabular}{@{}|c|rrr| @{}}
			\hline
			%& \multicolumn{3}{c}{R-1 chart} & \multicolumn{3}{c}{R-2 chart}\\
			%	\cmidrule(lr){2-4} \cmidrule(l){5-7}
			&		 	$\mu=1$  &	$\mu=2$		&	$\mu=3$ \\\hline
			$c=0$&	  {\bf25.22}	& 10.40 & 9.32 	 	\\
			$c=1$&	  52.85	& {\bf7.39}	 & 4.88	\\
			$c=2$& 	116.57 & 11.25& {\bf 3.70}	\\
			$c=3$& 	193.49 & 77.23 & 6.76	\\
			\hline
		\end{tabular}
 \end{table}

\begin{table}[!tb]
	\caption{ARL$_1$s of the R-2 control charts for $t$(4) and gamma(1,1).}
	\centering\label{t-2}
	\begin{tabular}{@{}|c|rrr| @{}}
		\hline
		&		 	$\mu=1$  &	$\mu=2$		&	$\mu=3$ \\\hline
		$t$(4)&	  23.27	&  5.16 & 3.11	 	\\
		gamma(1,1)&	 6.25 	& 3.84	 &2.81	\\
		\hline
	\end{tabular}

\end{table}

 \begin{table}[!htb]
 	\centering 
 	\caption{ARLs of the R-1 control charts and $c=2$.} \label{t-11}
 	\begin{tabular}{@{}|c|cccc|@{}} 
 		\hline
 		&$\mu=0$&		 	$\mu=1$  &	$\mu=2$		&	$\mu=3$ \\\hline
 		$r=6$	&200.61	 & 123.77	& 14.61 & 3.88 	 	\\
 		$r=8$	&210.89	 & 118.23	&     10.25 &  4.39 	 	\\         			%$r=10$		 &  87.8545	& 9.7356 &  5.0336 	 
 		\hline
 	\end{tabular}
 \end{table}

%The plot of probabilities of conditional scan statistics is given in Figure~\ref{f-1}  and the plot of probabilities of longest run statistics is given in Figure~\ref{f-2}. 
%\smallskip

\begin{table}[!bh]
	\caption{ARL$_1$s of the R-2 control charts for $\mathcal{N}$(0,1).}
	\begin{center}\label{T33}
		\begin{tabular}{@{}|c|rrr|@{}}
			\hline
			%& \multicolumn{3}{c}{R-1 chart} & \multicolumn{3}{c}{R-2 chart}\\
			%	\cmidrule(lr){2-4} \cmidrule(l){5-7}
			&		 	$m_0=30$  &	$m_0=50$		&	$m_0=100$ \\\hline
			$\mu=1$ &35.59  & 28.53 &  32.58\\
			$\mu=2$ &7.35 &   7.52  &  7.95\\
			$\mu=3$ &3.54 &   3.71  & 4.10\\
			\hline
		\end{tabular}
	\end{center}
\end{table}

%$\mu=1$ 35.5940   28.5270   32.5760
%$\mu=2$ 7.3450    7.5150    7.9490
%$\mu=3$ 3.5390    3.7090    4.1000

\vskip 3mm

\noindent 5. An Application

The proposed R-2 control charts are applied to the piston rings data in Tables 6.3 and 6E.7 in
\cite{Montgomery-2009}. \cite{Chakraborti-2009} used precedence statistics to analyze this same data.
The data in Table 6.3 consist of 25 samples. Each sample contains five inside diameter measurements of forged automobile engine piston rings. The $\bar{x}$ and $s$ control charts used in \cite{Montgomery-2009} indicate the process is in-control. Therefore, the 25 samples can be used to set up a phase II control chart. 

There are  additional 15 samples for the piston
ring process in Table 6E.7 in
\cite{Montgomery-2009}. To answer the question whether the process is still in-control for the 15 samples, the control chart is designed as follows.
%Our control charts allow  the monitoring of each single data point. Thus, we have 15$\times$5=75 points to be monitored using the longest run.   
To maintain a conditional probability of $\alpha = 1/400=0.0025$, 
the control chart is set up with  $ARL_0=400$. From the reference data, the summary statistics of $\bar{x} =74.001$ and of $\hat{\sigma}=0.01$ are used to help select the parameter $c$. To detect an upward mean shift,  the two-sigma threshold is used so $c = 74.001+2\times 0.01/\sqrt{5}=74.01$. The 15 sample means are then converted into a sequence of 0's and 1's, and this threshold results in 6 ones and 9 zeros.  Next, the charting procedure given in Algorithm 2 is implemented. The relevant statistics are given in Table~4. 
Note that controlling the conditional probability $\alpha$ exactly at $0.0025$ requires a  randomized test. Thus, according to Table~4, the proposed control chart is able to detect a signal as early as $t=9$ (see Figure~\ref{dataplot}) and no later than $t=14$.  To make it clear, the charting procedure is repeated 100 times, and on average the proposed control chart signals at $t\approx 11.86$.

\begin{table}[!ht] 	\label{4}
	\begin{center}
		\begin{tabular}{@{}|c|cccccccc|@{}}
			\hline
			$t$ & 1 & 2&3&4&5&6&7&8 \\\hline
			$\nu_n$ & 	0.0025	&0.0025	&0.0025	&0.0025	&0.0025	&0.0025	&0.0025	&0.0025	   	\\
			$k_n(\alpha)$ & 1& 1& 1& 1& 1& 1& 1& 1 \\
			$L_n$ & 0& 0& 0& 0& 0& 0& 0& 0  \\\hline
			$t$ & 	9	&	10	&11	&	12 & 13	&	14	&	15	& 	\\ \hline
			$\nu_n$ &0.0006	& 0.4288&0.0245 &0.0071 &0.0032 &0.0006&0.7125&	\\
			$k_n(\alpha)$	&	1	&2	&2&2&2&2&3&\\ 
			$L_n$ & 1&2&2&2&2&3&4 & \\\hline
		\end{tabular}
	\end{center}
	\caption{Summary statistics for the longest run rule control chart.}
\end{table}
\vskip 3mm

\noindent 6. Summary and Discussion

Distribution-free runs and patterns-type control charts are proposed. The proposed control charts can be applied to both continuous and discrete data. The method is illustrated using the longest run and scan statistics, but any other runs and patterns can be used as well; therefore, readers can use appropriate runs and patterns as monitoring statistics, which are capable of detecting variance shifts, to design distribution-free control charts.

To initiate the monitoring process, data are first converted into a Bernoulli sequence based on (\ref{111}). There are other transformations that can convert the data points into a Bernoulli sequence. For instance, the underlying continuous distribution $F_0$ can be used if it is known or can be accurately estimated. The transformed data then follow a uniform distribution, and a continuous scan statistic can monitor the mean or variance of the process. The proper parameters for $c$ and $r$ significantly improve performance.
 Some details are given in Section 4. 
The advantages of the control charts proposed here are three-fold: (i) they are truly distribution-free, and the  ARL$_0$ can be controlled  exactly at a pre-specified level, (ii) historical (reference) data are not necessary, and (iii) they are easy to understand and operate. The only assumption is that samples are i.i.d. from some unknown, underlying distribution $F_0$. 

However, there is a drawback to our proposed control charts.  
%Thus, if the process is out of control in the beginning, and there is no additional mean shift during the sampling process, the proposed control charts will not signal unless there is a subsequent larger mean shift.  
If the proposed control charts fail to detect the signal at an early stage, the performance of the charts will deteriorate. This is because whenever the control charts fail to detect a mean shift at time $t$, stronger evidence is required to trigger an alert due to the control charts having adapted to the shifted process after time $t$. There are several ways to mitigate this shortcoming. The user may restart the monitoring process or use a large reference sample. Moreover, a lagged version of the conditioning argument of the proposed charting procedure can be employed to prevent the control chart from adapting to the shifted process. In the adjusted procedures,
the charting statistics at time $t$ would depend on the ones at time $t-k$ instead of $t-1$, for some integer $1<k<t$. We will pursue this idea in a subsequent paper.

\vskip 3mm

 %\bibliographystyle{spbasic}      % mathematics and physical sciences
 %\bibliography{reference} 

\section*{Appendix}

	\subsection*{The Conditional Probabilities in (\ref{pro1}) and (\ref{ln})}
	
	The conditional probability in (\ref{pro1}) can be calculated by Theorem~\ref{thm1}. It follows from  Bayes' rule that 
	\begin{align*}
		&	P(S_n(r)< c_{n}(\alpha)|S_{n-1}(r)<c_{n-1}(\alpha),N_{n}=m)\\
		&\quad = \frac{P(S_n(r)< c_{n}(\alpha),S_{n-1}(r)<c_{n-1}(\alpha),N_{n}=m)}{P(S_{n-1}(r)<c_{n-1}(\alpha),N_{n}=m)}\\
		& \quad = \frac{P(S_n(r)< c_{n}(\alpha),S_{n-1}(r)<c_{n-1}(\alpha)|N_{n}=m)}{P(S_{n-1}(r)<c_{n-1}(\alpha)|N_{n}=m)}. 
	\end{align*}
	We have that 
	\begin{align*}
		& \mbox{if } c_{n}(\alpha)>c_{n-1}(\alpha), \mbox{ then }\\
		& \quad P(S_n(r)< c_{n}(\alpha)|S_{n-1}(r)<c_{n-1}(\alpha),N_{n}=m) = 1, \mbox{ and } \\
		& \mbox{if }  c_{n}(\alpha)\leq c_{n-1}(\alpha), \mbox{ then }\\
		&  \quad P(S_n(r)< c_{n}(\alpha),S_{n-1}(r)<c_{n-1}(\alpha)|N_{n}=m) = P(S_n(r)< c_{n}(\alpha)|N_{n}=m).
	\end{align*}
	Hence,
	\begin{align}
		&P(S_n(r)< c_{n}(\alpha)|S_{n-1}(r)<c_{n-1}(\alpha),N_{n}=m)\nonumber\\
		&\quad= \left\{\begin{array}{lll} \label{con1}
			1 & \mbox{if} & c_{n}(\alpha)>c_{n-1}(\alpha),\\
			\frac{P(S_n(r)< c_{n}(\alpha)|N_{n}=m)}{P(S_{n-1}(r)<c_{n-1}(\alpha)|N_{n}=m)} & \mbox{if} &  c_{n}(\alpha)\leq c_{n-1}(\alpha).\\
		\end{array}\right. 
	\end{align} 
	We  rewrite the probability in (\ref{con1})  as
	\begin{align*}
		&	P(S_{n-1}(r)<c_{n-1}(\alpha)|N_{n}=m)\\
		& \quad = 		 \frac{P(S_{n-1}(r)<c_{n-1}(\alpha),N_{n}=m)}{P(N_n = m)} \\
		%& \quad = \frac{P(S_{n-1}(r)<c_{n-1}(\alpha),N_{n-1}=m,X_n=0)+P(S_{n-1}(r)<c_{n-1}(\alpha),N_{n-1}=m-1,X_n = 1)}{P(N_n = m)}	\\
		%&\quad = \frac{P(S_{n-1}(r)<c_{n-1}(\alpha)|N_{n-1}=m)P(N_{n-1}=m,X_n=0)+P(S_{n-1}%(r)<c_{n-1}(\alpha)|N_{n-1}=m-1)P(N_{n-1}=m-1,X_n=1)}{P(N_n = m)}\\
		&\quad = \frac{P(N_{n-1}=m,X_n=0)}{P(N_n = m)}P(S_{n-1}(r)<c_{n-1}(\alpha)|N_{n-1}=m) \\
		&\qquad + \frac{P(N_{n-1}=m-1,X_n=1)}{P(N_n = m)}P(S_{n-1}(r)<c_{n-1}(\alpha)|N_{n-1}=m-1) \\
		&\quad = P(N_{n-1}=m,X_n=0|N_n = m)P(S_{n-1}(r)<c_{n-1}(\alpha)|N_{n-1}=m) \\
		&\qquad + P(N_{n-1}=m-1,X_n=1|N_n = m) P(S_{n-1}(r)<c_{n-1}(\alpha)|N_{n-1}=m-1),
	\end{align*}
	where $P(N_{n-1}=m,X_n=0|N_n = m) = {(n-m)}/{n}$ and $P(N_{n-1}=m-1,X_n=1|N_n = m) = {m}/{n}$.
	
	The conditional probability in  (\ref{ln}) is given by 
	\begin{align}
		&P(L_n(r)< k_{n}(\alpha)|L_{n-1}(r)<k_{n-1}(\alpha),N_{n}=m)\nonumber\\
		&\quad= \left\{\begin{array}{lll} 
			1 & \mbox{if} & k_{n}(\alpha)>k_{n-1}(\alpha),\\
			\frac{P(L_n(r)< k_{n}(\alpha)|N_{n}=m)}{P(L_{n-1}(r)<k_{n-1}(\alpha)|N_{n}=m)} & \mbox{if} &  k_{n}(\alpha)\leq k_{n-1}(\alpha).\\
		\end{array}\right. 
	\end{align} 
	
	\subsection*{Randomized Tests}
	To implement the charting procedures Section 3 describes, the main computational task is to determine the control limits $\{c_n(\alpha)\}$ and $\{k_n(\alpha)\}$ for the scan rule, and for the longest run rule control charts at a given $\alpha$ or ARL$_0$.  The longest run and scan statistics  are discrete; hence, to maintain an exact pre-specified $\alpha$ or ARL$_0$, randomized tests are used. We will show the detailed steps to determine $\{k_n(\alpha)\}$.  In a similar fashion, we can determine the control limits $\{c_n(\alpha)\}$.

	For a given $\alpha$, define a randomized test at time $n$ as 
	\begin{align}
		\phi_n  &= \left\{\begin{array}{lll} \label{test}
			1 & \mbox{if} \, L_n>k_n(\alpha),\\
			\nu_n & \mbox{if} \,    L_n=k_n(\alpha),\\
			0     &    \mbox{otherwise}.
		\end{array}\right. 
	\end{align}
	The probability in (\ref{ln}) becomes 
	\begin{align*}
		& P(\mbox{reject } H_0 \mbox{ at time } n| \mbox{ accept }  H_0 \mbox{ at time } n-1,N_n=m) \\
		&\quad = \frac{P(\mbox{reject } H_0 \mbox{ at time } n, \mbox{ accept }  H_0 \mbox{ at time } n-1|N_n=m)}{P(\mbox{ accept }  H_0 \mbox{ at time } n-1|N_n=m)}.
	\end{align*}
	By using (\ref{test}), we have 
	\begin{align*}
		& P(\mbox{reject } H_0 \mbox{ at time } n, \mbox{ accept }  H_0 \mbox{ at time } n-1|N_n=m)\\
		& \quad = E(\phi_n(1-\phi_{n-1})|N_n=m)\\
		& \quad = E(\phi_n|N_n=m)-E(\phi_n\phi_{n-1}|N_n=m),
	\end{align*}
	where $E(\phi_n|N_n=m) = P(L_n>k_n(\alpha)|N_n=m) + \nu_n P(L_n=k_n(\alpha)|N_n=m)$ and %$E(\phi_n\phi_{n-1}|N_n=m)$ is given by
	\begin{align*}
		E(\phi_n\phi_{n-1}|N_n=m)
		&  =  P(L_n>k_n(\alpha),L_{n-1}>k_{n-1}(\alpha)|N_n=m) \\
		& \qquad + \nu_n P(L_n=k_n(\alpha),L_{n-1}>k_{n-1}(\alpha)|N_n=m) \\
		& \qquad + \nu_{n-1} P(L_n>k_n(\alpha),L_{n-1}=k_{n-1}(\alpha)|N_n=m)\\
		& \qquad + \nu_n \nu_{n-1}P(L_n=k_n(\alpha),L_{n-1}=k_{n-1}(\alpha)|N_n=m).
	\end{align*}
	%Using a conditional argument and Theorem~\ref{thm1}, we can compute
	%\begin{align*}
	%& P(L_n=k_n(\alpha),L_{n-1}=k_{n-1}(\alpha)|N_n=m)\\&\quad = \frac{P(L_n=k_n(\alpha),L_{n-1}=k_{n-1}(\alpha),N_n=m)}{P(N_n=m)}\\&\quad=\frac{P(L_n=k_n(\alpha)|L_{n-1}=k_{n-1}(\alpha),N_n=m)P(L_{n-1}=k_{n-1}(\alpha),N_n=m)}{P(N_n=m)}\\&\quad = P(L_n=k_n(\alpha)|L_{n-1}=k_{n-1}(\alpha),N_n=m)P(L_{n-1}=k_{n-1}(\alpha)|N_n=m).
	%\end{align*}
	%To be specific, the probability $P(L_{n-1}=k_{n-1}(\alpha)|N_n=m)$ can be computed using Theorem~\ref{thm1}, and the probability $P(L_n=k_n(\alpha)|L_{n-1}=k_{n-1}(\alpha),N_n=m)$ can be computed by finding the one-step transition probability with the initial probability $P(L_{n-1}=k_{n-1}(\alpha)|N_n=m)$. 
	The value of $k_n(\alpha)$ can only be  $k_{n-1}(\alpha)$ or $k_{n-1}(\alpha)$+1. Hence, if $k_n(\alpha)=k_{n-1}(\alpha)=k$, then
	\begin{align*}
		&   P(L_n>k_n(\alpha),L_{n-1}>k_{n-1}(\alpha)|N_n=m)  = P(L_{n-1}>k|N_n=m) \\
		&   P(L_n=k_n(\alpha),L_{n-1}>k_{n-1}(\alpha)|N_n=m) = 0\\
		&   P(L_n>k_n(\alpha),L_{n-1}=k_{n-1}(\alpha)|N_n=m) = P(L_n>k,L_{n-1}>k-1|N_n=m)-P(L_{n-1}>k|N_n=m)\\
		&   P(L_n=k_n(\alpha),L_{n-1}=k_{n-1}(\alpha)|N_n=m)=P(L_n=k,L_{n-1}=k|N_n=m),
	\end{align*}
	and if $k_n(\alpha)=k_{n-1}(\alpha)+1=k+1$, then
	\begin{align*}
		&   P(L_n>k_n(\alpha),L_{n-1}>k_{n-1}(\alpha)|N_n=m)  = P(L_n>k+1,L_{n-1}>k|N_n=m) \\
		&   P(L_n=k_n(\alpha),L_{n-1}>k_{n-1}(\alpha)|N_n=m) = P(L_{n-1}>k|N_n=m) -P(L_n>k+1,L_{n-1}>k|N_n=m)\\
		&   P(L_n>k_n(\alpha),L_{n-1}=k_{n-1}(\alpha)|N_n=m) = 0 \\
		&   P(L_n=k_n(\alpha),L_{n-1}=k_{n-1}(\alpha)|N_n=m)=P(L_n=k+1,L_{n-1}=k|N_n=m).
	\end{align*}
	The above probabilities can be obtained by Theorem~\ref{thm1} with minor modifications.

\end{document}